\documentclass[twocolumn,showpacs,preprintnumbers,amsmath,amssymb,pre]{revtex4}

\usepackage[dvips]{graphicx}
\usepackage{dcolumn}
\usepackage{amssymb}
\usepackage{bm}

\begin{document}

\preprint{APS}

\title{Comment on ``Mechanical analog of temperature for the description of force distribution in static granular packings''}

\author{Philip T. Metzger}
\email{Philip.T.Metzger@nasa.gov}
\affiliation{%
The KSC Applied Physics Laboratory, The John F. Kennedy Space Center, NASA\\
YA-C3-E, Kennedy Space Center, Florida  32899
}%

\date{\today}

\begin{abstract}
It has been proposed by Ngan [Phys. Rev. E \textbf{68}, 011301 (2003)] that the granular contact force distribution may be analytically derived by minimizing the analog of a thermodynamic free energy, in this case consisting of the total potential energy stored in the compressed contacts minus a particular form of entropy weighted by a parameter.  The parameter is identified as a mechanical temperature.  I argue that the particular form of entropy cannot be correct and as a result the proposed method produces increasingly errant results for increasing grain rigidity.  This trend is evidenced in Ngan's published results and in other numerical simulations and experiments.
\end{abstract}

\maketitle

Ngan \cite{Ngan} recently proposed a functional minimization approach to derive the granular contact force probability distribution.  The method is different from the entropy maximization approaches that have been proposed by others because it explicitly accounts for the potential energy stored in the compressed granular contacts.  The previous methods had assumed Shannon's entropy for either the distribution $P_{x}$ of contact force Cartesian components $f_{x}$ (where $x$ is a principal stress axis) \cite{bagi}, 
\begin{equation}
S=-k\int_{0}^{\infty}\!\!\!\text{d}f_{x} \ P_{x}(f_{x})\ \ln P_{x}(f_{x}),
\end{equation}
or the distribution $P$ of contact force magnitudes $f$ \cite{kruytroth},
\begin{equation}
S=-k\int_{0}^{\infty}\!\!\!\text{d}f \ P(f)\ \ln P(f) .
\end{equation}
In either case, the entropy was maximized subject to the conservation of the number of contacts and the conservation of stress(es) in the packing.  This, along with other important variations of the methods, predicted a distribution function.  It is the latter of these two entropies that Ngan adopts.  In a paper now in preparation \cite{metzger} I show why both of these entropies are incorrect, because the density of states in the relevant phase space must be profoundly non-uniform, and it is this non-uniformity which is the source of the unique shape of the contact force distribution function in the region of weak forces.  In this Comment I will discuss the problem with the entropy as it is relevant to Ngan's hypothesis and show how the problem is revealed in the published results.

The functional proposed by Ngan is $F=U-\theta S$ where
\begin{equation}
U=\int_{0}^{\infty}\!\!\!\text{d}f \ P(f)\ W(f)\label{freeenergy}
\end{equation}
is the internal energy of the packing and where $W(f)$ is the work function for compression of a contact under normal force $f$.  Thus, $U$ is an analog of the Helmholtz free energy from thermodynamics and the parameter $\theta$ is the proposed mechanical analog of temperature.  This hypothesis produces interesting results because, for a Hertzian contact force law, the predicted distribution function closely fits the numerical simulation data both for 2D and, in the case of strong deformations, for 3D (at least to the statistical precision of the numerical data).  My own work has focused on a theoretical analysis of the case of perfect rigidity which is the limit in which Ngan's method is incorrect.  I am not able to comment on the region of large deformations, which I suspect is where Ngan's work makes its key contribution.  

\section{Gaussian versus Exponential}

In Ngan's hypothesis the form of the tail depends upon $W(f)$, predicting for Hertzian contacts a Gaussian tail $\sim \exp(-\beta f^{2})$ in 2D or a nearly Gaussian compressed exponential tail $\sim \exp(-\beta f^{5/3})$ in 3D.  This is in substantial agreement with statements in a paper by O'Hern, Langer, Liu, and Nagel (OLLN) \cite{ohern2}, which Ngan cites.  Before addressing the form of the entropy in Ngan's hypothesis, it is necessary to question whether Gaussian tails have really been observed in static granular force distributions.  I claim that they have not, except for cases with very small numbers of grains \cite{snoeijer} or large deformations \cite{makse, brujic}.  Five arguments make this case.

First, the tail normally does not appear exponential (a straight line on a semilogarithmic plot) until several multiples of the average value of force, $\left<f\right>$.  This onset is apparently dependent upon the dimensionality of the packing.  For example, Figs.~2 and~5 of Ref.~\onlinecite{radjaichaos} show the onset at $f \approx 3 \left<f\right>$ for 2D and $f \approx 2 \left<f\right>$ for 3D.  There may be other factors which hasten or delay the onset as well.  A number of the semilog plots in the published literature lack statistical precision for a sufficient range beyond the onset and thus may appear Gaussian.  This is because the eye tends to extrapolate the curvature it sees just prior to the loss of precision.  However, I have seen no case that is truly inconsistent with an exponential tail beyond the reasonable range of onsets except when there are large deformations \cite{makse, brujic} or a very small number of grains \cite{snoeijer}.  On the other hand, there are numerous examples which clearly show an exponential tail inconsistent with a Gaussian or other curved form.  These examples include 2D frictional simulations using contact dynamics (CD) and molecular dynamics \cite{radjaichaos, radjai}; 3D frictional simulations using CD \cite{radjaichaos}, a Hertzian contact law \cite{silbert}, and a Hookean contact law \cite{silbert, landry}; 3D frictionless simulations using two versions of the Lennard-Jones potential \cite{ohern1}; 3D frictional experiments \cite{experiment}; and 3D frictionless experiments with emulsions \cite{brujic}.

Second, the paper by OLLN claimed that 3D frictionless static packings with a harmonic potential produce a Gaussian tail \cite{ohern2}.  The thermal argument predicting this tail was not correctly applied to static packings.  The density of states of a thermal ensemble and the density of states of a static packing ensemble are organized by two mutually exclusive principles:  one by the conservation of energy and momentum with respect to time, the other by the conservation of forces with respect to several spatial dimensions.  In the transition as the temperature $T\to 0$, a packing does not become static until it locates one of the relatively rare locations in phase space that satisfy the highly organizing static equilibrium requirements.  There is no basis to assume that the density of states, or the potential energy distribution representing it, will have the same form after undergoing this self-organizing transition.

Third, the empirical data from OLLN's paper \cite{ohern2} do not contradict the existence of an exponential tail, either.  Their two Figs.~(3)(a) and~(3)(b) represent two types of ensemble averages.  In the first, an exponential tail onsetting at $f \approx 3 \left<f\right>$ fits arguably better than OLLN's proposed Gaussian over the same region \cite{footnote1}.  In the second, the tail is clearly exponential with an onset at $f \approx 2 \left<f\right>$.  For the harmonic potential in which $f= K x$, renormalizing the force scale of a packing is equivalent to renormalizing its spatial scale.  Hence, the first ensemble can be interpreted as all possible packing geometries compressed to achieve the same average force at the expense of different packing fractions.  The second ensemble is the set of all possible packing geometries compressed to achieve the same packing fraction but at the expense of different average forces.  Arguably it is the latter ensemble average, not the former, which represents the self-averaging of a much larger packing.  This is because locally averaged force fluctuations do occur in large packings (because the spatial distribution of forces is dominated by force chains and is therefore very heterogeneous), whereas the mean offset in OLLN's Fig.~(2)(b) implies that locally averaged packing fraction fluctuations become relatively small in an increasingly large packing.  Hence, this implies that an exponential onset in very large, frictionless, 2D packings ought to occur closer to $f \approx 2 \left<f\right>$, which is in agreement with the other data cited above.  However, regardless of which ensemble average is the ``correct'' one, both are at least consistent with an exponential tail.  I believe this conclusion is in better agreement with the statements found in the follow-on full-length paper by O'Hern, Silbert, Liu, and Nagel \cite{ohern3}.

Fourth, a simple argument can prove that, for frictionless packings in the limit of small deformations, the force distribution \textit{cannot} be a functional of $W(f)$.  Because a frictionless packing is isostatic \cite{isostatic}, all the forces can be derived deterministically from the geometry of the contact network and the imposed boundary loads, alone.  Therefore, the only role that $W(f)$ can play is by helping to decide what the geometry of the contact network shall be.  If $W(f)$ is relatively soft so that deformations are large, then the geometry of the contact network will be perturbed significantly, and then $W(f)$ may indeed affect the resulting forces.  However, in the limit when the deformations become vanishingly small, then the geometry of the contact network becomes independent of the form of $W(f)$ and the contact forces also become independent of the form of $W(f)$.  Therefore, the force distributions resulting from every $W(f)$ must approach a universal form in this limit.

Fifth, Bouchaud's argument \cite{bouchaud} with regard to the $q$ model persuasively explains the universal form of the tail.  He shows that the sufficient condition for an exponential tail is merely that some grains be allowed to tip all of their loads into one contact, which allows arbitrarily large forces to accumulate along particular force chains.  This argument applies to all cohesionless granular media regardless of $W(f)$, except for cases where the grains cannot freely tip.  We may draw an important inference:  when deformations become large, the formation of additional contacts increases the stability of the grains, hence allowing the forces to be more evenly distributed through space \cite{makse}.  This may erode the exponential tail so that it transits to a more rapidly decreasing form \cite{footnote2}.  

\section{The Role of the Entropy}

These arguments do not imply that Ngan's free-energy hypothesis is necessarily incorrect.  $W(f)$ is a measure of deformation, which is the relevant parameter when the tail becomes nonexponential.  The hypothesis does produce excellent results in the case of large deformations.  The tail of the predicted distribution also becomes steeper and more curved as the deformations increase, thus matching the observed trend \cite{makse, brujic}.  However, it can be seen in Ngan's 3D results, Fig.~(7), that the analytical predictions do not fit the simulation data for the 3D case with the least hydrostatic pressure applied to the packing, i.e., the case with the least grain deformation.  In fact, there is an unmistakable trend that with smaller pressures the predicted tail is increasingly distant from the simulation data which become increasingly consistent with an exponential tail \cite{footnote3}.  It is possible with just a small change in Ngan's hypothesis to make the predicted tail follow the trend toward an exponential.   This could be done by modifying the definition of the mechanical temperature $\theta$ so that $k$ remains constant and $\theta \propto \kappa^{-1}$.  Then, the influence of $W(f)$ would vanish where the isostatic argument says that it should, in the limit where deformations are small.

However, as shown in Ngan's Figs.~(1) and~(2), straightening the tail would raise a secondary but more fundamental problem because it would produce the incorrect features in the region of weak forces.  Then, because these features affect the average value of the distribution, the straightened tail would have the wrong exponential decay constant, $\beta=\left<f\right>$ instead of $\beta \approx 1.6\left<f\right>$.  Like the exponential tail, these weak-force features are so universal that they, too, are most likely the result of some fundamental, organizational behavior in granular media.  They are the small peak near the average value of force (or a plateau, or at least an abrupt change in slope in cases of severely anisotropic stress \cite{antony}) and the nonzero probability density at zero force.  

This secondary problem can be related to the Shannon's entropy maximization hypothesis of Kruyt and Rothenburg (KR) \cite{kruytroth}.  If Ngan's $U\to0$ with a nonvanishing $k$ as I have suggested, then the functional $F$ in Eq.~(\ref{freeenergy}) reduces to the entropy term alone.  Minimizing this is KR's method but with frictional grains.  If we took KR's prediction in the limit as the Coulomb coefficient of friction vanishes, $\mu\to0$, then it, too, becomes a purely exponential distribution, contrary to published results in the region of weak forces for frictionless packings \cite{Ngan, brujic, ohern1}.

The problem lies with Shannon's entropy because of its inherent implication that all possible sets of contact forces are equally probable, meaning that the density of states in the phase space with coordinates $\{f_{i}\mid i=1,\ldots,N\}$ is uniformly populated over the entire accessible region.  The accessible region is implied by the use of the Lagrange multipliers to be all states where the average force per contact is correct, or in the case of Ref.~\onlinecite{kruytroth} all states where the volumetrically averaged stresses match the externally applied stress tensor.  To be more accurate, the Shannon's entropy implies that any non-uniformity in the accessible region will be \textit{unbiased} with respect to the distribution of coordinates, so that the weaker forces are not left out by the nonuniformities any more often than are the stronger forces.  In granular packings this assumption is fundamentally incorrect because it neglects the most important organizing feature of granular packings:  the requirement that the grains be stable.  There is no analogous requirement in classical thermal systems, and it turns out that this requirement biases the density of states against weaker forces.  

An explanation of the bias begins by noting that any two contact forces on the same grain are strongly correlated, increasingly so as the contacts are further away from each other toward opposite sides of the grain \cite{silbert}.  For simplicity this Comment must illustrate how the correlation affects the density of states using only a special case, allowing the reader to draw the connections to the general case.
\begin{figure} 
\includegraphics[angle=0,width=0.3\textwidth]{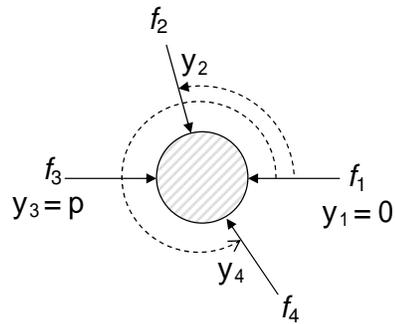}
\caption{\label{grainfig} Special case to illustrate why grains with below-average forces correspond to fewer stable locations in phase space than do grains with average forces.  }
\end{figure}
For an isotropic packing of frictionless 2D grains, consider one grain which has two of its contacts exactly opposite one another, $\psi_{1}=0$ and $\psi_{2}=\pi$ as shown in Fig.~\ref{grainfig}.  The force $f_{1}$ is clearly related to $f_{3}$, but not generally equal to it because of the contact forces $f_{2}$ and $f_{4}$ located in $\pi/3<\psi_{1}<2\pi/3$ and $4\pi/3<\psi_{4}<5\pi/3$, respectively.  Note that steric exclusion keeps the contacts arranged fairly predictably around a grain, which prevents the statistically averaged general case from deviating too far from this special case.  Static equilibrium requires
\begin{eqnarray}
f_{2} \sin\psi_{2} & = & -f_{4} \sin\psi_{4},\nonumber\\*
f_{1} & = & f_{3}-G,\label{stability}
\end{eqnarray}
where
\begin{equation}
G = f_{2}\cos\psi_{2}+f_{4}\cos\psi_{4}
\end{equation}
is the difference between $f_{1}$ and $f_{3}$.  Then $G = J^{-1} f_{4}$,
\begin{equation}
J^{-1}=\cos\psi_{4}-\cos\psi_{2}\frac{\sin\psi_{4}}{\sin\psi_{2}}.
\end{equation}
Neglecting that this is a special case (for illustrative simplicity), we assume that the distribution of $f_{4}$ is representative of all the forces in the packing.  These have a distribution $P_{f}(f)$, which we want to derive.  By changing variables the distribution of $G$ in terms of $P_{f}$ is
\begin{equation}
P_{G}(G)= \int_{\pi/3}^{2 \pi/3}\!\!\!\text{d}\psi_{2}\int_{4 \pi/3}^{5 \pi/3}\!\!\!\text{d}\psi_{4}\ |J|\ P_{f}\left(G\ J\right),\label{deltadist}
\end{equation}
where $J$ is identified as the Jacobian.  All we know about $P_{f}$ is that it is zero for negative arguments (tensile forces), but positive valued and normalized for positive arguments.  Assuming it is a continuous function its normalization implies that it has a bounded tail.  Over the range of integration, $J$ has odd symmetry in the sense that $J\left(\psi_{2},\psi_{4}\right)=-J\left(\pi-\psi_{2},3 \pi-\psi_{4}\right)$.  Hence, $P_{G}(G)$ must be an even function which is positive valued over $-\infty<G<\infty$, having a bounded tail in both extremes.  We cannot solve the distribution of $f_{1}$ by directly changing variables from $(G,f_{3})$ to $(f_{1},f_{3})$ because the pairs are not statistically independent and we do not know their joint probability distributions.  However, we know that the only stable configurations of this grain are those in which $G \le f_{3}$ because of Eq.~(\ref{stability}).  We can therefore calculate the proportions of the volume of phase space that include only the stable configurations of this grain with a particular value of $f_{3}$.  Since there are two equations of stability there are only two degrees of freedom in the force dimensions.  Integrating across only one of these to find the size of stable space as a function of the other, 
\begin{eqnarray}
V(f_{3}) & \propto & \int\!\!\!\int\!\!\text{d}^{2}\psi\int_{0}^{\infty}\!\!\!\text{d}f_{4}\ \ \Theta\left(f_{1}\right) P_{f}\left(f_{4}\right)\nonumber\\*
& \propto & \int_{-\infty}^{\infty}\!\!\!\text{d}G\ \ \Theta\left(f_{3} - G\right) P_{G}\left(G\right)\nonumber\\*
& \propto & \int_{-\infty}^{f_{3}}\!\!\!\text{d}G \ \ P_{G}\left(G\right),
\end{eqnarray}
where $\Theta$ is the Heaviside(unit step) function.  Hence, $V$ is a monotonically increasing function of $f_{3}$ which has a finite value at $f_{3}=0$.  This illustrates that the volume of the stable regions in phase space is biased against weaker forces, but is not vanishing for zero force.  Neglecting that this is a special case, the $P_{f}$ corresponding to the maximum volume of stable phase space is, therefore,
\begin{equation}
P_{f}(f)=e^{-\beta f} \int_{-\infty}^{f}\!\!\!\text{d}G \ \ P_{G}\left(G\right),
\end{equation}
which is a recursion equation in $P_{f}$ through Eq.~(\ref{deltadist}).  Thus, the relative slope of the exponential factor and the integral will determine the behavior of $P_{f}$ in the region of weak forces, and this provides the mechanism to explain variations in that region as a function of the anisotropy of the packing \cite{antony}.

Because of the bias against weak forces in stable phase space, the Shannon's entropy adopted by Ngan and others is not correct and this explains my claim why Ngan's predicted form for the case of vanishing deformations cannot fit both the tail and the region of weak forces for any choice of $\theta$ and $k$.  If we could define the entropy so that it accommodates the nonuniform density of states, then perhaps a particular choice of $\theta$ and $k$ can fit both the tail and the region of weak forces even in the low-deformation limit.  However, in the case of very deformable grains where Ngan's hypothesis works best, the principal role of the entropy term seems to be that it injects a logarithm into the equations.  Whether or not we will ever deduce a correct entropy maximization method for the case of perfect rigidity, it is safe to say that the actual form of the entropy with the relevant Lagrange multipliers will still contain a logarithm, so the applicability of Ngan's work in the large deformation limit is still an open and interesting question.

I am grateful to A.H.W. Ngan for the helpful interaction which has significantly improved this Comment.


\begin{thebibliography}{99}
\bibitem{footnote1}  Note that I am discussing only OLLN's case with the higher packing fraction.  This is the relevant case since it reduces the fine-tuned selection effect which biases the ensemble average.
\bibitem{footnote2} Apparently large deformations are not always sufficient to destroy the exponential tail \cite{erikson}.
\bibitem{footnote3}  Due to the limit of the statistical precision, Ngan's 2D data in the limit of low pressure are consistent with either an exponential tail or a Gaussian for $f>3 \left<f\right>$.
\end{thebibliography}
\end{document}